\documentclass[12pt]{article}
\usepackage[utf8]{inputenc}
\usepackage{graphicx} 
\usepackage{appendix}
\usepackage{amsmath}
\usepackage{mathtools}
\usepackage{longtable}
\usepackage{tabularx}
\usepackage{multirow}
\usepackage{authblk}
\title{Socio-demographic Determinants of Child Malnutrition Age 0-5 years in Bangladesh: NB and ZINB Approaches}
\author{Md Mehedi Hasan Bhuiyan}
\affil{Department of Statistics and Data science, University of Central Florida,USA}
\date{September 2023}

\makeatletter
\renewcommand{\@biblabel}[1]{\quad#1.}
\makeatother

\begin{document}

\maketitle

\begin{abstract}
Although child malnutrition is improving over the world in the last couple of decades, still now it is concerning issue among the developing countries including Bangladesh. In general, malnutrition is a dichotomous response variable fitted with logistic regression model. But in this study,  counting number of malnourished children in each household is defined as response variable. UNICEF with co-operating Bangladesh Bureau of Statistics (BBS) conducted Multiple Indicator Cluster Survey (MICS) covering 64000 households in Bangladesh by using two stage stratified sampling technique, where 21000 households have children age 0-5 years. We use bivariate analysis figuring out significant association  between  target and socio-demographic predictor variables. Then Negative binomial regression model is used over poisson regression model due to arising over-dispersion problem ($variance > mean$). Zero inflated negative binomial model also is applied for the excess of zeros in the target variable. Considering standard error and significant level of individual factors NB model provides better result as compare to ZINB.
\end{abstract}

\section{Introduction}
Malnutrition defined as deficiency of minerals, vitamins, and lack of physical need for growing up, consists of stunting, wasting, underweight and over-weight. Malnutrition in childhood  hampers linear growth of children in physically mentally through affecting communicable and non-communicable diseases, affects immune system and even tends to high risk of death,\cite{jubayer2022malnutrition,hossain2020malnutrition}. It also reduces the physical and mental potentiality in future, \cite{siddiqui2020intertwined}.\\

Child malnourishment is improving globally as well as developing countries including Bangladesh over the last decades. Stunted malnourished children age 0-5 years in the world were 33\% in 2000, and 22\% in 2022, whereas in perspective of Bangladesh, this rate was decreased 42\% in 2013, 28\% in 2019 respectively. In the case of wasting within age 0-5 years, the percentage of prevalent wasted  children over the world were reduced 8.7\% to 6.8\% from 2000 to 2022, while in Bangladesh, the percentage of the poor nourished children were 9.6\% in 2013 and 9.8\% in 2019. The global prevalent of underweight children were 25\% and 13\% in 1990,2019 respectively, other-hand, this percentage were 21.9\% and 16.6\% in 2013, 2019 respectively, in Bangladesh. The percentage of overweight children globally were increased slightly 5.3\% in 2000 and 5.6\% in 2022, These percentage in Bangladesh were 1.6\% in 2013 and 2.4\% \cite{mics2019, mics2013,world2023levels,akombi2017stunting,hossain2023prevalence}.\\

Child anthropometric measurement age 0-5 years is a continuous case. According to World health organization (WHO), the anthropometric measurement is converted into Z-score. For the child malnutritional status, Z-score lower than two standard deviation (SD) from the median of the reference population is defined as malnutrition. Malnutrition consists of underweight (weight for age, $<-2SD$), stunting (height for age, $<-2SD$), wasting (weight for height, $<-2SD$) and overweight Weight for height, $>2SD$, \cite{bloem20072006,de2012worldwide}.\\

Many studies use logit, probit models for the dichotomous response variable. On the other hand, poisson model is applied for counting response variable when mean equals to variance, while negative binomial (NB) is used for the over-disperson case (variance $>$ mean),\cite{islam2013predictors}. Moreover, zero inflated poisson model handles  counting  response variable, where excess zeros exist. But Zero inflated negative binomial (ZINB) works for over-disperson counting response variable with existing excess zeros, \cite{lee2023evaluation,lambert1992zero}.

\section{Methodology}
\subsection{Data Collection}
UNICEF with co-operating Bangladesh Bureau of Statistics (BBS) conducted Multiple Indicator Cluster Survey (MICS) in 2019 covering 64000 households in eight divisions by using two stage stratified sampling technique, where around 21000 households have children age 0-5 years.
\subsection{Variables selection}
Number of malnutrition is defined as total number  of children age 0-5 years in each malnourished categories in each household. In general, malnutrition is a dichotomous response variable fitted with logistic regression model. But in this study,  total counting number of malnourished children in each household is defined as response variable.\\
 
On the other hand, socio-demographic factors such as Gender (Male, Female), Wealth index (Low, Middle, High), Mother education (No education, Primary, Secondary, Higher), Father education (No education, Primary, Secondary, Higher), Residential area (Urban, Rural), Geographical location (known as Division) (Barishal, Chittagong, Dhaka, Khulna, Mymensingh, Rajshahi, Rangpur, Sylhet), Antenatal care during pregnancy (known as ANC visit) (Yes, No), delivery place (Home, Hospital), Child birth weight (Underweight, Average, Overweight),  Mother physical disability (Yes, No), Child illness (Yes, No), Taken antibiotic during illness (Yes, No), Number of child birth , household size define as independent variables.\\

\subsection{Method}

Probability mass function of negative binomial distribution (NB) is given by
$$
P(Y=y)=\frac{\Gamma(y+\tau)}{y ! \Gamma(\tau)}\left(\frac{\tau}{\lambda+\tau}\right)^\tau\left(\frac{\lambda}{\lambda+\tau}\right)^y, \quad y=0,1, \ldots ; \lambda, \tau>0
$$
where $\lambda=E(Y), \tau$ is a shape parameter, also known as over-dispersion, and $Y$ is the target variable. $Var(Y)$= $\lambda+\lambda^2 / \tau$. Clearly, the NB distribution tends to a Poisson distribution when $\tau \to \infty$. Consequently, the ZINB distribution is given by
$$
P(Y=y) \begin{cases}p+(1-p)\left(1+\frac{\lambda}{\tau}\right)^{-\tau}, & y=0 \\ (1-p) \frac{\Gamma(y+\tau)}{y ! \Gamma(\tau)}\left(1+\frac{\lambda}{\tau}\right)^{-\tau}\left(1+\frac{\tau}{\lambda}\right)^{-y}, & y=1,2, \ldots\end{cases}
$$

The mean and variance of the ZINB distribution are $E(Y)=(1-p) \lambda$ and $\operatorname{var}(Y)=$ $(1-p) \lambda(1+p \lambda+\lambda / \tau)$, respectively.\\

The ZINB regression model relates $p$ and $\lambda$ to covariates, that is,
$$
\log \left(\lambda_i\right)=x_i^{\prime} \beta \quad \text { and } \operatorname{logit}\left(p_i\right)=z_i^{\prime} \gamma, \quad(i=1, \ldots, n)
$$
where $x_i$ and $z_i$ are d- and q-dimensional vectors of covariates pertaining to the $i$ th subject, and with $\beta$ and $\gamma$ the corresponding vectors of regression coefficients, respectively. Log-likelihood of ZINB is given by
$$
\begin{aligned}
\mathcal{L}_z(\beta, \gamma, \tau ; \boldsymbol{\gamma}, X, Z)= & \sum_{i=1}^n \log \left(1+\mathrm{e}^{z_i^{\prime} \gamma}\right)-\sum_{i: y_i=0} \log \left(\mathrm{e}^{z_i^{\prime} \gamma}+\left(\frac{\mathrm{e}^{x_i^{\prime} \beta}+\tau}{\tau}\right)^{-\tau}\right) \\
& +\sum_{i: y_i>0}\left(\tau \log \left(\frac{\mathrm{e}^{x_i^{\prime} \beta}+\tau}{\tau}\right)+y_i \log \left(1+\mathrm{e}^{-x_i^{\prime} \beta} \tau\right)\right) \\
& +\sum_{i: y_i>0}\left(\log \Gamma(\tau)+\log \Gamma\left(1+y_i\right)-\log \Gamma\left(\tau+y_i\right)\right)
\end{aligned}
$$
where $X=\left(x_1, \ldots, x_n\right)$ and $Z=\left(z_1, \ldots, z_n\right)$ ,\cite{winkelmann2008econometric,weissbach2020consistency,lee2023evaluation,zeileis2008regression,mwalili2008zero}..


\section{Statistical analysis \& Result }
 We use bi-variate techniques such as cross-table, ANOVA, and chi-square test  figuring out significant association  between  target variable and socio-demographic predictor variables. Then NB regression model is used over poisson regression model due to arising over-dispersion problem $ var(Y): (1.003) > mean(Y): (0.701)$ with significant independent variables. We need to use ZINB regression model for excess of zeroes in the response variable, Figure 1 represent.\\

\begin{figure}
    \centering
    \caption{Frequency of child malnutrition age 0-5 years per household, Bangladesh}
    \includegraphics[width=16cm]{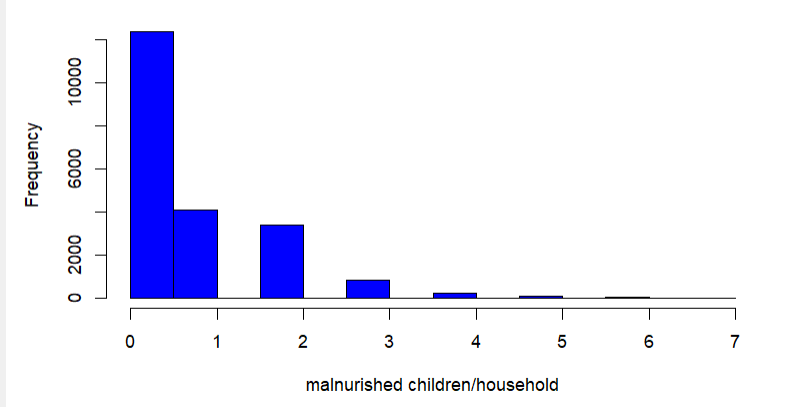}
    
\end{figure}

\begin{table}
\begin{center}
    \centering
    \caption {Contingency table for covariates with response variable}
    \begin{tabular}{|c|c|c|} 
    \hline
    Characteristics & $\chi^{2}$-value & P-value  \\\hline 
    Gender & $19.077^{***}$ & 0.007 \\ \hline
    Area & $46.127^{***}$ & 0.000 \\ \hline
    Antenatal care (ANC) & $200.68^{***}$ & 0.000 \\ \hline
    Delivery place & $365.56^{***}$ & 0.000 \\ \hline
    Mother education & $513.2^{***}$ & 0.000 \\ \hline
    Father education & $260.81^{***}$&0.000 \\ \hline
    Child weight at birth & $157.217^{***}$ & 0.000 \\ \hline
    Wealth index &$349.07^{***}$&0.000\\\hline
    Geographical location &$396.67^{***}$&0.000\\\hline
    Number of child birth & $543.7^{***}$ &0.000\\\hline
    Mother physical disability & $3.6941$ &0.345\\\hline
    Household size & $149.01^{***}$ &0.000\\\hline
    Child illness & $20.695^{***}$ &0.0042\\\hline
    Taken antibiotic during illness & $18.004^{**}$ &0.01195\\\hline
    \end{tabular}\\
      \emph{Note:} '*' 10\%, '**' 5\%, '***' 1\% significant level.
\end{center}
    
\end{table}

\begin{table}
 \begin{center}
    \centering
    \caption{Coefficents of covariate from NB and ZINB regression models}
    \begin{tabular}{|p{3.5cm}|p{2.2cm}|p{2.8cm}|p{2.8cm}|p{2.8cm}|}
      \hline
      \textbf{Variables } & \textbf{variable labels} & \textbf{IRR of NB} & \textbf{IRR of ZINB (link:log)} & \textbf{IRR of ZINB (link:logit)}\\
      \hline
      \multirow{2}{*}{Antenatal care (ANC)} & Yes & $0.840^{***}(0.039)$ &  $0.906^{***}(0.042) $& $1.278^{**}(0.120)$\\ 
       & No (ref) & - & -&-\\ \hline
        \multirow{2}{*}{Delivery place} & Home & $ 1.162^{***}(0.035)$  & $1.149^{***}(0.041) $ & $0.967(0.097)$\\ 
       & Hospital (ref)  & - & -&-\\ \hline
        \multirow{2}{*}{Residential area} & Rural & $0.954(0.042) $ & $1.087(0.053)$ & $1.390^{**}(0.129)$\\ 
       & Urban (ref)  & - & -&-\\ \hline
         \multirow{3}{*}{Wealth quintile} & Middle & $0.874^{***}(0.043) $  & $0.956(0.048) $& $1.284^{**}( 0.113)$\\ 
         &Rich& $0.822^{***}(0.044)$ & $0.962(0.054) $ & $1.502^{***}(0.122)$\\
       & poor (ref)  & - & -&-\\ \hline
         \multirow{2}{*}{Gender} & Female & $0.872^{***}(0.030)$ & $0.997(0.034)$ & $1.448^{***}(0.083)$\\ 
       & Male (ref)  & - & -&-\\ \hline  
        \multirow{4}{*}{Mother education} & Primary & $ 0.909^{*}(0.054) $ & $0.943(0.054)$& $1.215(0.172)$ \\ 
         &Secondary & $ 0.860^{***}(0.056) $ & $ 0.931(0.059) $ & $ 1.380^{*}(0.176) $\\
         &Higher& $0.887(0.073)$ & $0.936(0.084)$ & $1.306(0.218)$\\
       & None  (ref)  & - & -&-\\ \hline

        \multirow{4}{*}{Father education} & Primary & $1.020(0.038)$  & $1.050(0.042)$ & $1.120(0.110)$\\ 
         &Secondary&$0.915^{**}(0.043)$& $1.015(0.050)$&$1.332^{**}(0.118)$\\
         &Higher& $0.832^{***}(0.065)$&$0.938(0.081)$&$1.355^{*}(0.172)$\\
       & None  (ref)  & - & -&-\\ \hline
       
        \multirow{5}{*}{Geographical location} & Chittagong & $1.240^{***}(0.039)$  & $1.060$& $0.637^{**}(0.175)$\\ 
         &Dhaka&$1.146^{**}(0.062)$&$1.097(0.072)$&$0.888(0.174)$\\
         &Rangpur&$1.179^{**}(0.066)$&$1.077(0.078)$&$0.787(0.183)$\\
         &Sylhet&$1.633^{***}(0.069)$&$1.397^{***}(0.075)$&$0.634^{**}(0.188)$\\
       & Barishal(ref) &-&-&-\\ \hline
        \multirow{3}{*}{Child birth weight} & average & $1.177^{***}(0.052)$  & $1.040(0.065)$& $0.749^{**}(0.128)$\\ 
         &Underweight&$1.662^{***}(0.058)$&$1.284^{***}(0.070)$&$0.500^{***}(0.153)$\\
       & Overweight  (ref)  & - & -&-\\ \hline
       Total No. of child birth& & $1.101^{***}(0.013)$& $1.085^{***}(0.013)$&$0.963(0.035)$\\\hline  
    \end{tabular}
    
  \end{center}
   \emph{Note:} '*' 10\%, '**' 5\%, '***' 1\%, inside (): standard error of coefficient
\end{table}

Table 1 represents that summary of bivariate analysis between response and feature variables. All the predictor variables have significant association with the target variable except the variable: mother physical disability.\\
Table 2 shows statistical result of NB and ZINB. The table only represents statistically  associated variables under 10\% level of significant. NB regression model provides good result with respect to the standard error of coefficient of co-variates, Children suffering in malnutrition for mothers having antenatal care during pregnancy are 16\% lower as compare to the children whose mothers not having antenatal care. The  incidence rate of child malnourishment is higher (16.2\%) for  delivery in home than delivery in hospital. The percentages of children having poor health condition are lower for the middle and rich class family (13\% and 18\% respectively) compare to poor class family. Female children are leading poor health condition is $13\%$ lower than male children. Malnourished children of primary educated mothers and secondary educated mothers are 9\% and 14\% lower compare to children of  none educated mothers. In the case of father education, the incidence rate of malnourished children for secondary and higher educated father are 0.92 times and 0.83 times lower compare to the children of none educated fathers. Considering geographical location, the percentages of children with malnourishment in Chittagong, Dhaka, Rangpur and Sylhet  are 24\%, 15\%, 18\% and 63.3\% more compare to the children of Barishal location. The Children of average weight during birth is having 18\% more malnourished compare to the children of overweight at birth, this percentages is much higher for the children of underweight at birth (66\%). The incidence rate of malnourished children is increasing 10\% more with increasing children numbers.

\section{Discussion}
Mother antenatal care (ANC visit) is positively associated with decreasing child malnutrition. This result is consistent with other study, \cite{siddiqi2011malnutrition,hossain2020malnutrition}. In this study, the child malnutrition is not significantly associated with urban and rural area. But other many studies show that living area is significantly associated with child malnutrition, \cite{islam2020reducing,hossain2023prevalence}. It is widely known that place of delivery is link with child health condition.  Home delivery can cause some infections with children that may hamper short term or long term linear growth of children. The same scenario is found in this study. Children are living in Chittagong, Dhaka, Rangpur, and Sylhet are leading poor health status. This is consistent result with other studies, \cite{rahman2021investigate,islam2020reducing}. \\
Many studies show that parents education is directly associated with child malnutrition. Maternal education is more focusing on reducing the malnutrition. The increasing parents education  level improves child health conditions. In this study, parents education level has significantly associated with reducing  child malnutrition status. This result is consistent with others studies, \cite{siddiqi2011malnutrition,hossain2020malnutrition,hossain2023prevalence,bbaale2014maternal}.\\
Economic inequality is a big burden for improving child poor health condition. Poor family can not maintain daily standard consumption for the family. Several studies noted child poor health status inversely associated with family economic index that means improving economic level decreases child malnutrition. This result also consistent with this study, \cite{delpeuch2000economic,hossain2020malnutrition,larrea2005does}.\\
Gender discrimination is prevalence through out the world. Many studies show male children are getting more favors as compare to female children. On the other hand, Th probability of being malnourished among female children are lower than male children,\cite{mehrotra2006child,choudhury2000gender,khan2015gender}. But our study favors female children. Male children are more malnourished as compare to female children. \\
Its is commonly known that child nourishment is strongly associated with child birth weight. Study also validates low birth weight is a factor of child malnourishment. Children of low birth weight are more vulnerable to infection with several communicable and none-communicable diseases such as Diarrhea, abnormal heart bit, respiratory problem, etc.. This also hampers children growth during childhood. It is also counted as a risk factor for mortality and morbidity.\cite{hossain2020malnutrition,ntenda2019association}. Our study also shows  a significant positive association exists between low birth weight children and malnutrition . 

\section{Conclusion}
The study represents that NB model is a good model figuring out socio-demographic factors affecting child malnutrition in Bangladesh.
ANC visit, wealth index, parents education have significant positive impact on reducing child malnutrition. The authority of Bangladesh can launch free medical service for the pregnant women. Government can also provide more subsidiary in education sector for the improving the education level in urban and rural area for both gender. As a developing country, economic inequality is another burden for the child malnutrition in Bangladesh. Taking a new economic policy considering geographical locations could have long time impact on improving the nutritional status of children in household level. Average number of child birth is higher in Bangladesh. We need to discourage to have more children through advertisement or proper policy.

\section*{Acknowledgements}
We acknowledge UNICEF and BBS for getting free data set 'Multiple Indicator Cluster Survey-2019, Bangladesh'.

 \bibliographystyle{plain}
 \bibliography{references}

\end{document}